\documentstyle [aaspp,epsf]{article}
\tighten
\def \tablerule {\noalign {\hrule}} 

\begin{document}

\lefthead {Sahni, Sathyaprakash and Shandarin}
\righthead {Shapefinders}

\title {Shapefinders: a new shape
diagnostic for large scale structure.}
\author {Varun Sahni$^{\rm a}$, B.S. Sathyaprakash$^{\rm b}$, 
Sergei F. Shandarin$^{\rm c}$}
\affil {$^{\rm a}$Inter-University Centre for Astronomy \& Astrophysics,
Post Bag 4, Pune 411007, India}
\affil {$^{\rm b}$Department of Physics and Astronomy, UWCC, Cardiff, 
CF2 3YB, U.K.}
\affil {$^{\rm c}$Department of Physics and Astronomy, University of Kansas, 
Lawrence, KS 66045}

\begin {abstract}

We construct a set of shape-finders which determine shapes of compact
surfaces (iso-density surfaces in galaxy surveys or N-body simulations)
without fitting them to ellipsoidal configurations as done earlier.
The new indicators arise from simple, geometrical considerations 
and are derived from fundamental properties of a surface such as its
volume, surface area,
integrated mean curvature and connectivity characterized by
the Genus. 
These `Shapefinders' could
be used to diagnose the presence of filaments, pancakes and ribbons
in large scale structure. Their lower-dimensional generalization
may be useful for the study of
two-dimensional distributions such as temperature maps of the Cosmic
Microwave Background.

\end {abstract}


The large scale structure of the Universe is remarkably rich in visual
texture. At different density thresholds the clustering pattern
has been variously described as `meatball-like',
`sponge-like', `bubble-like', `network of surfaces' etc. 
(\cite{zesh82,mel90,delgh91}).
Attempts to quantify this pattern, in red-shift surveys
of galaxies and in N-body simulations, have been made using topological
discriminators such as the genus curve and percolation statistics 
(\cite{zel82,sh83,gmd86}; 
for a recent discussion see \cite{sss97}),
and also by applying minimal spanning trees and 
statistics sensitive to 
`shape' (\cite{bs92,lv95,sss96,dave}, for a review see \cite{sc95}).

Recently Minkowski functionals have been used to 
characterize the {\it global} geometrical and topological properties of 
pointwise distributions, e.g. galaxy or cluster catalogues (\cite{mbw94}). 
In this paper
we suggest to use a set of geometrical parameters derived from Minkowski
functionals to describe the geometry and topology of {\it individual} objects
such as superclusters of galaxies. Minkowski functionals have a very convenient 
property of additivity and therefore they can be used for both 
an isolated structure (such as a cluster or supercluster of galaxies) or 
for a group of structures (such as all structures in an N-body simulation
or a galaxy catalog). Probably using the Minkowski
functionals themselves is the most general and universal approach to quantitatively representing
the geometry
of superclusters and voids, but often we are interested in the characteristic 
dimensions of an object as well. We have therefore devised {\it Shapefinders}
-- statistics 
having dimensions of [Length] as well as dimensionless statistics
to characterize the morphology
of large scale structure.

The morphology of superclusters and voids is 
likely to differ for different scenarios of structure formation
and a study of supercluster-void shapes could help
distinguish between radically different alternatives such as
gravitational
instability, seed models of structure formation and models based on
explosions or
`mini-bangs'.

It is well known that systems evolving under gravitational
instability percolate at higher density thresholds
corresponding to progressively lower values of the filling factor, 
which suggests that structures
in the percolating phase are more likely to be sheet- or filament-like
since sheets and filaments occupy a larger surface area than a sphere at
a given
volume, and therefore percolate more easily (\cite{ks93,sss96})
For a fluid that has evolved as a result of  gravitational instability, a low 
filling factor at percolation is also suggested by the Zeldovich
approximation which predicts that the first singularities to
form are pancake-like (\cite{zel82,sz89}). 
However, it is unlikely that these pancakes will be
strictly planar objects. Instead it is more natural that they will resemble, 
in a manner
of speaking, the curved two-dimensional surface of a cup 
(\cite{ashz82,shetal95}).
Recent work suggests that soon after pancake formation,
the density distribution becomes dominated by filaments which act as
bridges connecting neighboring clusters (\cite{bkf96,sss96}), 
with pancakes remaining statistically significant\footnote {It is 
worth stressing the difference between formation of pancakes, filaments,
or clumps and topological properties of density fields. 
According to the definition of Zeldovich, pancakes form in regions of 
three stream-flows or between shock fronts. 
Other authors however label any flattened region as a pancake. 
In any {\it generic} density field regions having densities 
above a  sufficiently high thresholds look like compact regions (all three
dimensions are similar). At a lower threshold $\rho_1$ the connected 
structure forms in the higher density regions $\rho > \rho_1$,
the lower density regions $\rho < \rho_1$
remain connected (in other words both percolate). Then, at even lower
threshold $\rho_2 < \rho_1$ the lower density regions $\rho < \rho_2$
cease percolating. The second percolation transition is sometimes associated 
with  the formation of pancakes (e.g. \cite{bkf96}).}.

Most galaxy catalogues reveal structures with typical scales $\sim
50 $ Mpc., some such as the Great Wall, appear to be even bigger.
The present finite size of surveys, together with the fact that most
of them are limited to surveying galaxies within a wedge shaped region, 
prevents us from establishing
whether the visual structures we see are truly filamentary (one-dimensional)
or they appear filamentary because the geometry of the survey
prevents us from acquiring a fully three-dimensional perspective
(filaments in a wedge type survey could, for instance, be slices
of two-dimensional `sheets'). Upcoming large red-shift surveys such as the
2dF survey at the Anglo-Australian Telescope and the Sloan Digital Sky
Survey promise to reveal large scale structures in their full glory and
shed more light on their three-dimensional shapes.

The importance of trying to quantify shapes of clusters
and superclusters, in galaxy surveys and in simulations has, in recent 
years, led to a discussion of different statistical tools which may be
sensitive to `shape'. While such statistics have had a measured amount
of success, it is fair to say that none of them is entirely satisfactory.
A central feature of some shape indicators is that they
describe the shape of
a collection of points (equivalently -- an overdense region)
by evaluating its moment of inertia tensor, which is similar to 
fitting by an ellipsoid. The ratios of the principal axes then provide
a means of ascertaining whether the structure is oblate or prolate.
This method has been widely used in determining the luminosity profiles
of galaxies 
and remains a powerful tool for classifying the
projected shapes of ellipticals. 
Its efficacy as a discriminator for large scale structure is, however, not 
quite as
obvious. 
Whereas general physical principles  suggest that the shapes
of galaxies
should be predominantly elliptical or spiral, the shape of 
overdense
regions in large scale structure 
(clusters and superclusters of galaxies) is likely to be far
more complicated and less prone to a classification in terms of `eikonal'
shapes such as ellipsoids. Indeed, results of N-body simulations 
show that, when viewed at different density thresholds, shapes of compact
surfaces can vary
widely, ranging from approximately ellipsoidal (at high densities),
to topologically complicated `spongy' shapes 
at moderate density thresholds.

An example of a multiply-connected surface often seen at 
moderate thresholds in simulations
is a torus. Clearly a statistic which attempts to describe the shape of
a torus by fitting with an ellipsoid would be widely off the mark 
since it would lead us to
conclude that the torus has a pronounced oblate shape and would miss
completely its tubular form -- which is more like a
one-dimensional filament.
A tendency to model shapes using pre-defined
`eikonal' forms, can lead to an exaggerated emphasis of oblateness or
sphericity over
filamentarity and could easily bias our understanding of the morphology
of large scale structure (\cite{sssf97}).
In this paper we introduce new `shape finders' which 
are free from the above drawbacks, and probe the shape
of an object 
without any preordained reference
to an eikonal shape.

A compact surface, which could be an isodensity contour above or below
a given density threshold in an N-body simulation, 
can be characterized by the following four 
quantities\footnote{These are
also known as Minkowski functionals and have been applied to Cosmology
by \cite{mbw94}.}:
(i) Volume $V$, (ii) surface area $S$, (iii) integrated mean curvature:
$C = {1\over 2}\int(\kappa_1 + \kappa_2) dS$, 
(iv) integrated Gaussian curvature (genus):
${\cal G} = -{1\over 4\pi}\int \kappa_1\kappa_2 dS$,
where $\kappa_1 \equiv 1/R_1$ and 
$\kappa_2 \equiv 1/R_2$ are the principal curvatures of the surface.
Multiply-connected surfaces have ${\cal G} \ge 0$ while simply connected
have ${\cal G} < 0$. 

We introduce three {\it Shapefinders} each having dimensions of [Length]: 
${\cal H}_i, i = 1, 2, 3$ where
${\cal H}_1 = V/S$, ${\cal H}_2 = S/C$ and ${\cal H}_3 = C$
(for multiply-connected surfaces $C/{\cal G}$ may be more appropriate than
$C$).
Taken together, the trio ${\cal H}_i$ provide 
robust
and convenient measures of `shape' as will be shown below.
Based on ${\cal H}$ we can also
define a pair of very useful dimensionless Shapefinders:
${\bf{\cal K}} \equiv ({\cal K}_1, {\cal K}_2)$ where 
\begin{equation}
{\cal K}_1 = \frac{{\cal H}_2 - {\cal H}_1}{{\cal H}_2 + {\cal H}_1},\ \
{\cal K}_2 = \frac{{\cal H}_3 - {\cal H}_2}{{\cal H}_3 + {\cal H}_2}.
\end {equation}
We note that  ${\cal K}_{1,2} \le 1$ by construction.
$\bf{{\cal K}}$ can be regarded as a two-dimensional vector
whose amplitude and direction determine the shape 
of an arbitrary  three-dimensional surface.
Combined with the genus, we get the dimensionless triad
$({\cal K}_1, {\cal K}_2, {\cal G})$ giving information about shape as well as
topology. 
The Shapefinders ${\cal H}_i$, having dimensions of {\it length},
can be thought of as describing the 
spatial dimensions of an object. Thus,
an ideal pancake (having vanishing thickness but not necessarily planar) 
has one characteristic dimension much smaller than the remaining two,
so that ${\cal H}_1 \ll {\cal H}_2 \simeq
{\cal H}_3$ and ${\bf{\cal K}} \simeq (1, 0)$. An
ideal filament (a one-dimensional object but not necessarily straight) 
has two characteristic dimensions much smaller than the third so that
 ${\cal H}_1 \simeq {\cal H}_2 \ll {\cal H}_3$ and
${\bf{\cal K}} \simeq (0, 1)$. 
All three dimensions of a sphere are equal resulting in
${\cal H}_1 \simeq {\cal H}_2 \simeq {\cal H}_3$ and
${\bf{\cal K}} \simeq (0, 0)$. In addition, an interesting surface to 
consider is a `ribbon', for which ${\cal H}_1 \ll {\cal H}_2 \ll {\cal H}_3$
and ${\bf{\cal K}} \simeq (1, 1)$.

The genus ${\cal G}$, 
integrated mean curvature $C$, and surface area $S$, can be
derived for an arbitrary surface by considering its first and second fundamental
forms.
Consider a local coordinate patch on a surface of 
class $\ge C^2$, ${\bf r} \equiv \bf{r}(\theta,\phi)$,
then the {\it first} ({\bf I})  and {\it second} ({\bf II})
fundamental forms of the surface are:
\begin{eqnarray}
{\bf I} & = ~{\bf dr}\cdot {\bf dr} & = ~E ~d\theta^2 + 2F ~d\theta ~d\phi + G ~d\phi^2\nonumber\\
{\bf II} & = ~- {\bf dr}\cdot {\bf dn} & = ~L ~d\theta^2 + 2M ~d\theta d\phi + N d\phi^2
\label{eq:3a}
\end{eqnarray}
where ${\bf n}$ is the unit normal to the surface 
${\bf n} = {\bf r}_{\phi}\times{\bf r}_{\theta}/\vert {\bf
r}_{\phi}\times{\bf r}_{\theta}\vert,$ and
\begin{eqnarray}
& E = {\bf r}_{\theta}\cdot{\bf r}_{\theta},~
G = {\bf r}_{\phi}\cdot{\bf r}_{\phi},~
F = {\bf r}_{\phi}\cdot{\bf r}_{\theta} & \nonumber\\
& L = {\bf r}_{\theta\theta}\cdot{\bf n},~
N = {\bf r}_{\phi\phi}\cdot{\bf n},~
M = {\bf r}_{\phi\theta}\cdot{\bf n}, &
\label{eq:3}
\end{eqnarray}
${\bf r}_{\phi} \equiv
\partial{\bf r}/\partial\phi,$ ${\bf r}_{\phi\phi} \equiv
\partial^2{\bf r}/\partial\phi^2,$ etc.
As a result one gets (see e.g. \cite{difgeom})
\begin{eqnarray}
S & = &\int\int \sqrt{EG - F^2}~d\phi d\theta\nonumber\\
\kappa_1 + \kappa_2 & = &{EN + GL - 2FM\over EG - F^2},\nonumber\\
\kappa_1\kappa_2 & = &{LN - M^2\over EG - F^2}\nonumber\\
C & = &\int\int\frac{\kappa_1 + \kappa_2}{2} dS, \nonumber\\
{\cal G} & =&- \frac{1}{4\pi}\int\int\kappa_1\kappa_2 dS.
\label{eq:4} 
\end{eqnarray}
We elaborate on the shape statistics by applying them to two
surfaces -- an ellipsoid and a torus. 
Consider first
the triaxial ellipsoid having volume $V = \frac{4\pi}{3}abc$ and parametric form
\begin{equation}
{\bf{r}} = a(\sin{\theta}\cos{\phi})\hat{x} +
b(\sin{\theta}\sin{\phi})\hat{y}
+ c(\cos{\theta})\hat{z}
\end{equation}
where $0 \le \phi \le 2\pi$, $0 \le \theta \le \pi$. 

Results for the trio of dimensionful Shapefinders ${\cal H}_i \equiv $
($V/S$, $S/C$, $C$)
and the dimensionless Shapefinders ${\bf{\cal K}} \equiv ({\cal K}_1,
{\cal K}_2)$ are shown
in Tables \ref{table1} \& \ref{table2} for the triaxial ellipsoid with axis 
($a, b, c$)
and its deformations into
a pancake, filament and sphere respectively. 
In all cases the Shapefinders
have
been normalized to give ${\cal H}_i = R~$ $({\cal K}_i = 0$) for a sphere of
radius $R$.

Our results for idealized `eikonal' surfaces are as follows:
pancakes ($a, b=a, c$), $ a \gg c$, ${\bf{\cal K}} \simeq (1, 0)$;
filaments $(a, b, c=b)$, $ a \gg b$, ${\bf{\cal K} }\simeq (0, 1)$;
ribbons $(a, b, c)$, $ a \gg b \gg c $, ${\bf{\cal K}} \simeq (1, 1)$;
spheres ($a, b=a, c=a$), ${\bf{\cal K}} \simeq (0, 0)$.
Thus, the deformation of a sphere into a pancake or filament results
in the accompanying increase in ${\bf{\cal K}}$: $(0, 0) \rightarrow (1,
0)$ (pancake), or  
$(0, 0) \rightarrow (0, 1)$ (filament).
While deforming a filament into a pancake (and vice versa) it is
possible to 
encounter an intermediate 
surface which resembles a ribbon, for this surface
 ${\bf{\cal K}} \simeq (\alpha, \alpha)$, 
$\alpha \le 1$.
Tables \ref{table1} \& \ref{table2} also show an interesting
symmetry between oblate and prolate 
ellipsoids. Consider an oblate ellipsoid ($a, a, c$), with 
${\bf{\cal K}} = ({\cal K}_1, {\cal K}_2)$, then, for the prolate 
ellipsoid
$(a, c, c)$, ${\bf{\cal K}} = ({\cal K}_2, {\cal K}_1)$!
For shapes close to spherical, both ${\cal K}_1, {\cal K}_2$ are small, and
it is more appropriate to study the ratio ${\cal K}_i/{\cal K}_j$ to assess
departure from sphericity. Thus for pancakes ${\cal K}_1/{\cal K}_2 > 1$,
filaments ${\cal K}_2/{\cal K}_1 > 1$ and for 
ribbons ${\cal K}_1/{\cal K}_2 \simeq 1$. In addition the statistic
${\cal K}_{1,2}^p$, $p < 1$, will also accentuate small departures from
sphericity.

Next we consider a torus with elliptical cross-section
having the parametric form
\begin{equation}
{\bf{r}} = (b + c\sin{\phi})\cos{\theta}~\hat{x} +
(b + c\sin{\phi})\sin{\theta}~\hat{y} + a(\cos{\phi})\hat{z}
\end{equation}
where $a,c < b$, $0 \le \phi, \theta < 2\pi$.
The toroidal tube with diameter $2\pi b$ has an elliptical cross-section, 
$a$ \& $c$ being respectively radii of curvature 
perpendicular and parallel to the plane of the
torus. The usual circular torus is given by $a=c$.

Table \ref{table3} shows ${\cal K}$ \&
${\cal H}$ for a torus with an elliptical cross-section which is an
interesting surface since, like the 
triaxial ellipsoid, it has three sets of numbers defining its shape but
it also has a hole which makes it topologically non-trivial. 
Such surfaces (with holes) generically arise in large
scale structure surveys and in N-body simulations, especially at moderate
density thresholds.
Our results for the torus
are qualitatively similar to those for the ellipsoid but with interesting 
new 
features. 
For instance one has two possibilities for a ribbon: Ribbon1 resembles a 
finger-ring, whereas Ribbon2 is a disc-with-large-hole, see Fig.1.
(In the extreme case $c \simeq b$, Ribbon2 becomes like Pancake2,
and, for
$a \simeq b$, Ribbon1 becomes like Pancake1.)

Finally, for both the ellipsoid and the torus, ${\bf{\cal K}}$
remains invariant under a similarity transformation:
($a, b, c) \rightarrow (\alpha a, \alpha b, \alpha c$). 

It is quite remarkable that ${\bf {\cal H}}$ and ${\bf{\cal K}}$ 
give compatible 
results for similar shapes obtained by deformating two very
dissimilar bodies, a torus and an ellipsoid.
We feel this indicates that the Shapefinders
introduced by us are robust
and can be used for identifying topologically complicated shapes.  
In a companion paper we shall apply the shape-statistic to
N-body simulations (\cite{sss97a})\footnote{It is interesting that  other Shapefinders
besides those mentioned above,
can be constructed
out of
the four Minkowski functionals $V, S, C$ and ${\cal G}$.
For instance the dimensionless pair ${\bf\Delta}\equiv (\Delta_1, \Delta_2)$, 
where
$\Delta_1 = S^3/V^2 - 1$, $\Delta_2 = (C^2/S)^3 - 1$, is also a good
probe of pancakes, filaments and ribbons. 
${\bf \Delta}$ shares many of the features of ${\bf{\cal K}}$ (the main
difference being that, unlike ${\bf {\cal K}}$,
${\bf\Delta}$ is not bounded from above $\Delta_{1,2} \ge 0$).
If like ${\bf {\cal K}}$, we normalise ${\bf \Delta}$ so that 
$\Delta_{1,2} = 0$ for spheres, then, for (i) pancakes $\Delta_1 > \Delta_2$,
(ii) filaments $\Delta_2 > \Delta_1$, (iii) ribbons $\Delta_1 \simeq
\Delta_2 \neq 0$.}.
As emphasised by \cite{sss96}, shape diagnosis combined with
percolation theory provide a powerful tool with
which to study gravitational clustering. Percolation theory demonstrates
that the number of clusters/voids in a simulation always peaks at a
threshold just above/below percolation. 
Thus, it makes sense to study shapes of individual clusters/voids at this
threshold since the number of distinct isolated surfaces (boundaries of
clusters/voids) is largest.
In addition, it is likely that most surfaces at such low thresholds will
be multiply-connected with a large genus, perhaps making it 
appropriate to define the third shape statistic to be $C/{\cal G}$ instead of
$C$ as we have been assuming so far.

The statistic suggested by us could also be used to study more general shapes
than those appearing in large scale structure. For instance one could
use them to study the shapes of concentrated cosmic magnetic fields 
which might have important astrophysical consequences (\cite{rkb97}).
Finally, the two-dimensional 
Shapefinders ${\cal H}_1 = S/L$, ${\cal H}_2 = L$ and ${\cal K} = ({\cal H}_2
- {\cal H}_1)/({\cal H}_2 + {\cal H}_1)$
 ($L$ is the circumference of a curve bounding
a two-dimensional area $S$), combined with the two-dimensional genus, 
could prove useful when studying shapes and topologies of two-dimensional 
contours defining `hot and cold spots' in the Cosmic Microwave
Background, or isodensity surfaces in projection data. (Values of 
${\cal K}$ range from zero for a circle to unity for a filament.)

\noindent {\bf Acknowledgments:}
We acknowledge stimulating discussions with Sanjeev Dhurandhar and Somak
Raychaudhury. SFS acknowledges financial support from the NSF-EPSCoR 
program and NASA grant NAG5-4039.

\clearpage

\begin {figure}[ht]
   \begin{center}
         \epsfxsize 5 true in \epsfbox {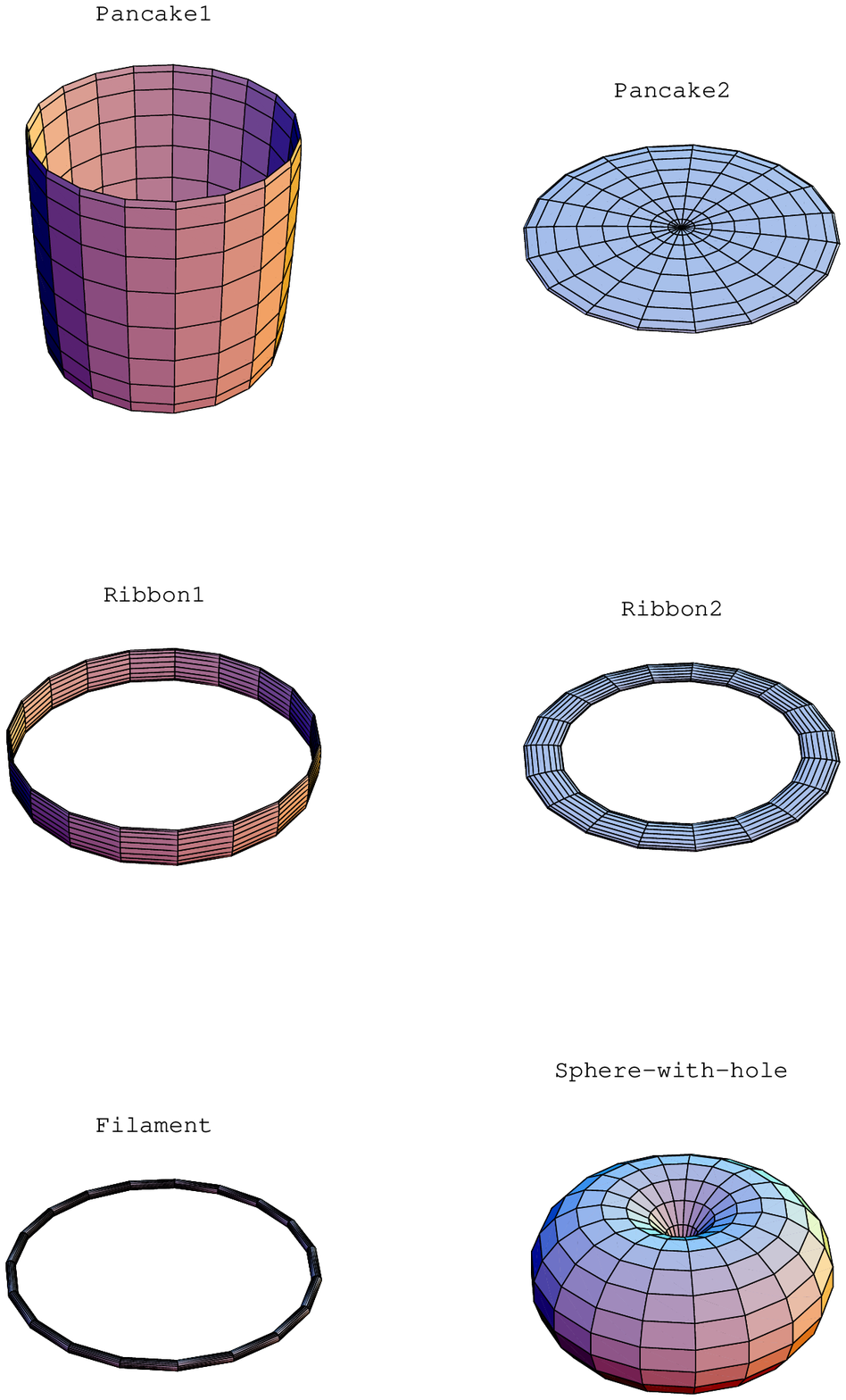}
   \caption {\small Deformations of a torus with an elliptical cross section.}
   \end{center}
\end {figure}

\clearpage

\begin {table} [t]
\caption {Shapefinders for a triaxial ellipsoid with axis $a,b,c$.
($V/S$, $S/C$, $C$  have dimensions of length, 
${\cal K}_{1,2}$ are dimensionless.)
The second column (M) is a description of the  morphology of the object
based on its dimensions: P= Pancake, F= Filament, R= Ribbon, 
S= Sphere.}

\vskip 0.2 true cm
\label{table1}
\begin {tabular} {cccccc}
\tablerule
$a,b,c$ & M  & $({\cal K}_1, {\cal K}_2)$ &$V/S$ & $S/C$ & $C$\\ 
\tablerule 
(100, 100, 3  ) & P & ($0.83, ~0.10$) & 5.98 & 63.9 & 78.6 \\ 
(100, 3, 3)     & F & ($0.10, ~0.83$) & 3.82 & 4.70 & 50.2 \\ 
(200, 20, 2)    & R & ($0.67, ~0.67$) & 3.94 & 20.0 & 101  \\ 
(100, 100, 100) & S & ($0.99, ~0.00$) & 100  & 100  & 100  \\
\tablerule
\end {tabular}
\end {table}
\begin {table}[t]
\caption {Deformations of a triaxial ellipsoid with axis $a,b,c$. }
\label{table2}
\vskip 0.2 true cm 
\begin {tabular} {cc} 
\tablerule 
$a,b,c$ & $({\cal K}_1, {\cal K}_2)$\\ 
\tablerule 
\multicolumn {2}{c}{Sphere $\rightarrow$ Filament}\\
\tablerule 
(100, 100, 100)&  ($0.000, ~0.000$)\\
(100, 80, 80)  &  ($0.004, ~0.005$)\\
(100, 50, 50)  &  ($0.028, ~0.054$)\\
(100, 20, 20)  &  ($0.077, ~0.300$)\\
(100, 10, 10)  &  ($0.095, ~0.540$)\\
(100, 3, 3)    &  ($0.100, ~0.830$)\\    
\tablerule
\multicolumn {2}{c}{Sphere $\rightarrow$ Pancake}\\
\tablerule 
(100, 100, 100)& ($0.000, ~0.000$)\\
(100, 100, 80) & ($0.005, ~0.004$)\\
(100, 100, 50) & ($0.054, ~0.028$)\\
(100, 100, 20) & ($0.300, ~0.077$)\\ 
(100, 100, 10) & ($0.540, ~0.095$)\\ 
(100, 100, 3)  & ($0.830, ~0.100$)\\ 
\tablerule 
\multicolumn {2}{c}{Pancake $\rightarrow$ Filament}\\
\tablerule 
(100, 100, 3) & ($0.830, ~0.100$)\\
(100, 70, 3)  & ($0.800, ~0.130$) \\ 
(100, 30, 3)  & ($0.650, ~0.330$) \\ 
(100, 10, 3)  & ($0.330, ~0.650$) \\ 
(100, 3, 3)   & ($0.100, ~0.830$) \\ 
\tablerule 
\end {tabular} 
\end {table} 
\begin {table} [ht]
\caption {Shapefinders for a torus of radius $b$ 
having an elliptical cross-section with axis
$a,c$, 
($b > a,c$).  ($V/S$, $S/C$, $C$ have dimensions of length, 
${\cal K}_{1,2}$ are dimensionless.) 
The second column (M) is a description of the morphology of the
object based on its dimensions: P1,P2=Pancake, 
F=Filament, R1,R2=Ribbon, S=Sphere-with-hole.}
\vskip 0.2 true cm 
\label{table3}
\begin {tabular} {cccccc} 
\tablerule 
$b,a,c$ & M & $({\cal K}_1, {\cal K}_2)$ & $V/S$ & $S/C$ & $C$ \\  
\tablerule  
(100, 99, 3) & P1 & ($0.90, ~0.03$) & 7.05 & 136    & 144  \\ 
(100, 3, 99) & P2 & ($0.88, ~0.20$) & 7.05 & 114    & 173  \\
(100, 3, 3)  & F  & ($0.14, ~0.93$) & 4.5  &   6.00 & 157  \\  
(150, 20, 2) & R1 & ($0.70, ~0.80$) & 4.64 &  25.9  & 235  \\
(150, 2, 20) & R2 & ($0.70, ~0.80$) & 4.64 &  25.9  & 235  \\
(20, 19, 19) & S  & ($0.14, ~-0.09$) & 28.5 & 38.0  &  31.4\\ 
\tablerule 
\end {tabular} 
\end {table}

\end {document}